# A Study on Library Resources with Services Satisfaction based on Library Users Affiliated Colleges to Solapur University

*Patel Adam Burhansab\*, Dr. M Sadik Batcha\*\* and Muneer Ahmad\*\*\**

\* Research Scholar, Department of Library and Information Science, Annamalai University, Annamalai Nagar, Tamil nadu
\*\* Research Supervisor and Mentor, Professor and Librarian, Annamalai University, Annamalai Nagar, Tamil nadu
\*\*\* Research Scholar, Department of Library and Information Science, Annamalai University, Annamalai Nagar, Tamil nadu

## Abstract

*The main aim of this study was to assess and evaluate user satisfaction with library resources and services among library users associated with Solapur University. The current research shows the level of users' satisfaction with different library resources and services offered by college libraries. The research found that a vast number of respondents were pleased with library facilities and services. The research is designed to achieve users' satisfaction in the library to investigate the level of satisfaction towards library resources and services with regards to 26 colleges of Solapur University based in Maharashtra. Information in the form of data has been collected from colleges and on the basis of users results; analysis needs to analyze users' satisfaction.*

## 1. Introduction

College libraries are hubs of academic life for improving the student's knowledge. It is expensive to explore the level of user satisfaction through the services of the College Library. New students go to college every year with different needs and expectations. With the advancement of technological advancements and the array of data that is becoming accessible to users, the aggressive pressure for academic libraries will continue to increase. The performance of the administration of the library can only be measured by the extent at which its administration and services are used. The three notable foundations in colleges are laboratories, instructors/classrooms, and libraries containing rich and updated resources and services including hardware that can support the work of teaching, learning, and research. The purpose of this review was to assess the administration provided by the institution and the level of users' satisfaction among students and faculties. Different types of resources and services are provided in academic libraries in view of user pre-requisites. Academic libraries are also expected to know about the needs of users and fulfill their demands and needs. The fundamental goals of this section are to differentiate between respondents from different categories of users like undergraduate, postgraduate students and faculty members to learn about the usage of library resources and services.

## 2. Review of Literature

In India and abroad, there are several studies conducted on users' satisfaction level. The authors tried to include a few important studies. **Chavez et al. (2005)** have studied the general users of the Paradise Valley Community College Library were satisfied with the resources, facilities and circulation services and suggested a repeat survey in two years to compare student satisfaction levels. **Uganneya and ldachaba (2005)** jointly conducted a study and found that circulation and information technology facilities at 7 University of Agricultural

Libraries was inadequate. The level of library use was found to be poor and also the library did not fulfill the users' knowledge needs properly. **Sriram and Rajev (2014)** researched to classify the different resources and facilities needed by Sur University College Sultanate of Oman academic library users and their degree of effect on the satisfaction of their users. The usage of library facilities, the satisfaction of users with library resources and services, and the actions of students and research scholars finding information at Tezpur University were examined by **Saikia and Gohain (2013)**. The analysis showed that in addressing the multidimensional demands of students and scholarly academics for knowledge and skills, the library plays a vital role. Usage guidance is believed to be important to assist library users in fulfilling their demands for information and to inform users of the library's available tools and services.

**Poll and Payne (2006)** jointly found in a study that the benefits of using library resources by academic library users can be measured in terms of information obtained, information literacy, educational and professional success, social addition, and amplifying the well-being of the entity.

**Sowole (1995)** carried out study users satisfaction level of library is accomplished by offering the information tools and services needed. **Nadozie (2006)** Analysis has shown that the facilities needed to provide qualitative services to the library are either accessible or not completely accessible. The use of library facilities, and definitely their satisfaction with library services, was discussed by **Abagai (1993)** based on the availability of trained staff, information resources, and library accommodation. **Nwalo (1997)** performed a review with the assistance of library facilities and facilities requirements and regulations; the word lib evaluation. **Cullen and Calvert (1993)** found that the assessment of inputs based on collection, budget, staff resources and measures of process competence are markers of the understanding of library users of the resources and services provided by the library. **(Burhansab, Batcha & Ahmad, 2020)** studied the use of electronic resources/information by library users at selected college libraries affiliated to Solapur

University. Well-structured questionnaires were used to gather data from 1022 library users from selected colleges affiliated to Solapur University. The result showed that the preponderance of users from Aided (33.51%), Self-financing (26.10%) and educational colleges (43.24%) preferred to visit the Library once in three days.

## 3. Objectives

- To identify the services provided by the library and services preferred by users.

- To assess the opinion about library resources, facilities and services.

- To measure user satisfaction levels with library resources and services in college libraries.

## 4. Research Methodology

### 4.1. Sampling and Questionnaire Design

A survey of 26 colleges including Aided colleges, Self-financing colleges, Engineering colleges, Educational colleges and Pharmacy colleges affiliated to Solapur University based in Maharashtra, India were selected under the study. There was a structured questionnaire designed. The questionnaire consisted of different questions in the following categories: demographics, e-resource use, e-resource access, skills and training, e-resource challenges and benefits, among others. Two open-ended questions were also asked by respondents about the implement and enhancement of e-resources in the libraries of their colleges. A total of 1350 questionnaires were distributed to users of the library of selected colleges under study, out of which 1022 were received back. The response rate of received questionnaire was 75.70 per cent. The data analysis was performed using the Social Sciences Statistical Package (SPSS Version 17).

### 4.2. Sample Selection

Sample is selected on the basis of confidence level 99% and confidence interval 5%. Following are the samples of the study of various colleges affiliated to Solapur University, Solapur. The sample and total population taken for the study are shown in Tabulation (Table 4.3)

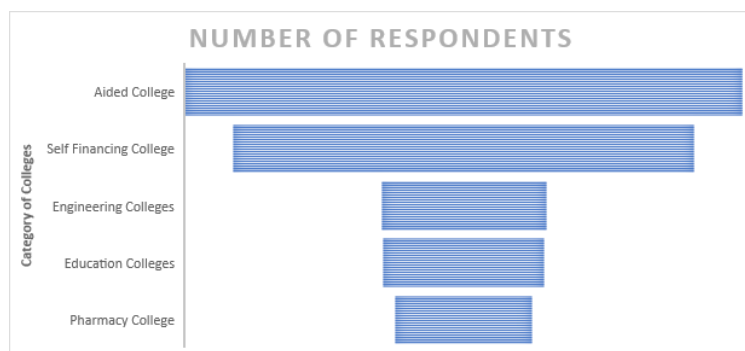

**Table 4.3: Representation of Respondents from Different Categories of colleges**

| S. No | Category of the Colleges | Male | Female | Total |
|---|---|---|---|---|
| | Aided College | 245 | 140 | 385 |
| | Self-Financing College | 174 | 144 | 318 |
| | Engineering Colleges | 67 | 47 | 114 |
| | Education Colleges | 66 | 45 | 111 |
| | Pharmacy College | 59 | 35 | 94 |
| | **Total** | **611** | **411** | **1022** |

## 5. Results and Discussion

### 5.1: Frequency of Using ICT Resources

**Table 5.1 Frequency of Using ICT Resources**

| ICT Resources / Frequency | Library online catalogue | | E-Journals | | Electronic Thesis & Dissertation | | E-Books | | Bibliographic Databases | | Printing | | Scanning | | Institutional Repository | |
|---|---|---|---|---|---|---|---|---|---|---|---|---|---|---|---|---|
| Daily | 122 | 11.9 % | 153 | 15.0 % | 111 | 10.9 % | 160 | 15.7 % | 94 | 9.2 % | 178 | 17.4 % | 106 | 10.4 % | 99 | 9.7 % |
| Weekly | 282 | 27.6 % | 395 | 38.6 % | 245 | 24.0 % | 326 | 31.9 % | 232 | 22.7 % | 318 | 31.1 % | 255 | 25.0 % | 178 | 17.4 % |
| Monthly | 355 | 34.7 % | 273 | 26.7 % | 292 | 28.6 % | 308 | 30.1 % | 415 | 40.6 % | 369 | 30.1 % | 326 | 31.9 % | 283 | 27.7 % |
| Never | 263 | 25.7 % | 201 | 19.7 % | 374 | 36.6 % | 228 | 22.3 % | 281 | 27.5 % | 157 | 22.3 % | 335 | 32.8 % | 462 | 45.2 % |
| Total | 1022 | 100 % | 1022 | 100 % | 1022 | 100 % | 1022 | 100 % | 1022 | 100 % | 1022 | 100 % | 1022 | 100 % | 1022 | 100 % |
| Ranking | 6 | Monthly | 1 | Weekly | 7 | Never | 2 | Weekly | 4 | Monthly | 5 | Monthly | 3 | Never | 8 | Never |

The above table explains the frequency of using ICT resources. While ranking the usage of ICT resource, majority of the users are found using E-Journals (38.6%) and E-books (31.9%) on weekly basis. Mainstream of users from various college libraries are using Bibliographic databases (40.7%), printing (36.2%), and Library on line catalogue (34.8%) on monthly basis. However, some of the ICT Resources like E-thesis and dissertation (36.6%), Scanning (32.7%) and Institutional Repository (45.2%) are found at the scale of never by the most of the users.

### 5.2 OPAC Usage satisfaction in the Libraries

**Table 5.2: OPAC Usage satisfaction in the Libraries (College-wise)**

| Colleges / OPAC | Aided Colleges | | Self-Finance College | | Engineering Colleges | | Education Colleges | | Pharmacy Colleges | | Total | |
|---|---|---|---|---|---|---|---|---|---|---|---|---|
| Highly Dissatisfied | 46 | 11.9 % | 22 | 6.9 % | 8 | 7.0 % | 10 | 9.0 % | 22 | 23.4 % | 108 | 10.6 % |
| Dissatisfied | 46 | 11.9 % | 37 | 11.6 % | 19 | 16.7 % | 12 | 10.8 % | 22 | 23.4 % | 136 | 13.3 % |
| Satisfied | 191 | 49.6 % | 211 | 66.4 % | 69 | 60.5 % | 64 | 57.7 % | 34 | 36.2 % | 569 | 55.7 % |
| Highly Satisfied | 55 | 14.3 % | 25 | 7.9 % | 11 | 9.6 % | 17 | 15.3 % | 10 | 10.6 % | 118 | 11.5 % |
| Neutral | 47 | 12.2 % | 23 | 7.2 % | 7 | 6.1 % | 8 | 7.2 % | 6 | 6.4 % | 91 | 8.9 % |
| Total | 385 | 100 % | 318 | 100 % | 114 | 100 % | 111 | 100 % | 94 | 100 % | 1022 | 100 % |
| Chi Square = 155.3 P Value = 0.000 | | | | | | | | | | | | |

It is explained from the Table 5.2 that majority of the users from Self Financing College (66.4%), Engineering College (60.5%), Education College (57.7%), Aided College (49.6%), and Pharmacy College (36.2%) are found satisfied with the present OPAC system. On the other hand, 23.4% of users from Pharmacy College are observed highly dissatisfied and in addition another 23.4% of the users have opted dissatisfied with the OPAC system. A Higher number of users from Aided Colleges (16.9%), self-financing colleges (11.6%), Engineering college (16.9%), Education College (10.8%) and Pharmacy College (23.4%) are opting the point scale of dissatisfied with the OPAC system. The calculated Chi square value is 155.3 and p value zero is found less than 0.01 indicates that there is a significant difference among the users in various libraries regarding the level of satisfaction on OPAC.

Greater part of the users (55.7%) from entire libraries claimed 'satisfied' and another 11.6% of users are 'highly satisfied' with the current OPAC facilities.

Moreover, in the highly dissatisfied (10.5%), dissatisfied (13.3%) and never satisfied (8.9%) groups cannot be ignored. OPAC is the bibliography to the library as well as to the users. Thus, it can be concluded that usage of OPAC in Libraries are almost fulfilled, yet the number of dissatisfied scale users are not less in numbers. The libraries in which satisfaction level is resulted very low, the improved OPAC facilities would be made more accessible to the users. It may be due to by a group of users not aware of its usage. They should be given training in using OPAC. Hence the libraries should try to overcome these problems so that extreme utilization of OPAC can be possible.

### 5.3: Library Website usage in the Libraries

The table brings out the result that majority of the users opted searching of study materials at a higher response (34.5%). They are benefited with study materials through the college website as the foremost preference followed by announcements and events (23.8%) to know about the various programs conducted by Libraries. Working hours (20.9%) and membership information (13.6%) are also searched by the user and selected in the third and fourth option respectively. Useful links to resources (28.2%) is selected by the users in the Fifth position.

### Table 5.3: Library Websites Usage in Libraries

| Purpose | Preferences | | | | | | | | | | | Total |
|---------|---|---|---|---|---|---|---|---|---|---|---|---|
| | 1 | | 2 | | 3 | | 4 | | 5 | | 6 | |
| Working hours | 163 | 15.9 % | 158 | 15.5 % | 185 | 18.1 % | 175 | 17.1 % | 167 | 16.3 % | 174 | 17.0 % | 1022 |
| Membership Information | 151 | 14.8 % | 153 | 15.0 % | 160 | 15.7 % | 221 | 21.6 % | 191 | 18.7 % | 146 | 14.3 % | 1022 |
| Announcement and Events | 152 | 14.9 % | 212 | 20.7 % | 195 | 19.1 % | 160 | 15.7 % | 151 | 14.8 % | 152 | 14.9 % | 1022 |
| Study Material | 231 | 22.6 % | 196 | 19.2 % | 152 | 14.9 % | 152 | 14.9 % | 144 | 14.1 % | 147 | 14.4 % | 1022 |
| Useful links to resources | 187 | 18.3 % | 166 | 16.2 % | 160 | 15.7 % | 144 | 14.1 % | 210 | 20.5 % | 155 | 15.2 % | 1022 |
| On line Catalogue (Web OAPC) | 144 | 14.1 % | 145 | 14.2 % | 151 | 14.8 % | 180 | 17.6 % | 152 | 14.9 % | 250 | 24.5 % | 1022 |

Considering the overall analysis of the website usage of the total Libraries, it is shown that proper guidance should be given to the users by the authorities for using the website. It helps to use the various Library resources and spontaneously increase the usage. It is common factor that only a limited number of users are using the Library Website effectively and efficiently for the academic purpose. Conducting user awareness programs by the Libraries concerning the website resources will help to explicit the usage of resources.

**5.4: Reasons for using IT in the Libraries (College-wise)**

**5.4: Reasons for Using IT in the Libraries**

| Colleges /Reason for using IT | Aided Colleges | | Self-Finance College | | Engineering Colleges | | Education Colleges | | Pharmacy Colleges | | Total | |
|---|---|---|---|---|---|---|---|---|---|---|---|---|
| Time saving | 72 | 18.7 % | 62 | 19.5 % | 15 | 13.2 % | 27 | 24.3 % | 17 | 18.1 % | 193 | 18.9 % |
| Easier | 104 | 27.0 % | 95 | 29.9 % | 38 | 33.3 % | 22 | 19.8 % | 18 | 19.1 % | 277 | 27.1 % |
| Faster | 78 | 20.3 % | 53 | 16.7 % | 26 | 22.8 % | 32 | 28.8 % | 31 | 33.0 % | 220 | 21.5 % |
| Wider | 63 | 16.4 % | 58 | 18.2 % | 19 | 16.7 % | 15 | 13.5 % | 14 | 14.9 % | 169 | 16.5 % |
| Current | 68 | 17.7 % | 50 | 15.7 % | 16 | 14.0 % | 15 | 13.5 % | 14 | 14.9 % | 163 | 15.9 % |
| Total | 385 | 100 % | 318 | 100 % | 114 | 100 % | 111 | 100 % | 94 | 100 % | 1022 | 100 % |
| Chi Square = 91.2 P Value = 0.000 | | | | | | | | | | | | |

Table 5.4 shows that the principal choice of using IT said by the users of Aided Colleges (27.0%), Self-Finance College (29.9%) Engineering college (33.3%) is easier access of information but in the case of Education College (28.8%) and Pharmacy College (33.0%) the choice of IT differs as faster access. At the same time self-financing college and Education colleges prefer time saving as the other reason. A small group of users from pharmacy college (19.2%) selects IT for easier access of information.

Overall information reveals that among various reasons for using IT Services, easier (27.1%) is ranked one. As IT services provide with fastest (21.5%) technologies most of the user rank it to number two. Comparatively a good number of users also support IT for its time saving process (18.9%). However, in the grading of wider access (16.5%) and current information (16%) there is a slight difference given by the users. Chi square value 91.2% and p-value 0.000 shows that there exists significant variation in opinion regarding the reasons for using various IT resources and services.

Moreover, there are no doubts that all the users find IT services as easier, current, Faster and time saving than manual methods. Thus, it can be seen that IT services are inevitable part of information seekers. It is easier to access information, time saving Faster etc. Still manual services have its own significance in this fast-running technology world. However, the modern technologies play an enthusiastic role to provide information within the stipulated time.

**5.5 Satisfaction level of working Hours of Libraries**

The working hours of Library and time required for using the Library among the user should be satisfactory in the present situation. Regarding the working hours of the libraries some of the college Libraries in India are providing services 24 x 7 basis. Now the commencement of Digital Library through Wi-Fi network campus helps the users for speedy and round the clock retrieval of the resources.

**Table 5.5: Working Hours Satisfaction in the Libraries (College-Wise)**

| Colleges /Satisfaction of Working hours | Aided Colleges | | Self-Finance College | | Engineering Colleges | | Education Colleges | | Pharmacy Colleges | | Total | |
|---|---|---|---|---|---|---|---|---|---|---|---|---|
| Highly Dissatisfied | 48 | 12.5 % | 40 | 12.6 % | 11 | 9.6 % | 7 | 6.3 % | 19 | 20.2 % | 125 | 12.2 % |
| Dissatisfied | 47 | 12.2 % | 42 | 13.2 % | 15 | 13.2 % | 12 | 10.8 % | 10 | 10.6 % | 126 | 12.3 % |
| Satisfied | 183 | 47.5 % | 142 | 44.7 % | 72 | 63.2 % | 67 | 60.4 % | 52 | 55.3 % | 516 | 50.5 % |
| Highly Satisfied | 58 | 15.1 % | 52 | 16.4 % | 9 | 7.9 % | 18 | 16.2 % | 7 | 7.4 % | 144 | 14.1 % |
| Neutral | 49 | 12.7 % | 42 | 13.2 % | 7 | 6.1 % | 7 | 6.3 % | 6 | 6.4 % | 111 | 10.9 % |
| Total | 385 | 100 % | 318 | 100 % | 114 | 100 % | 111 | 100 % | 94 | 100 % | 1022 | 100 % |
| Chi Square = 71.9 P Value = 0.000 | | | | | | | | | | | | |

The table 5.5 shows level of satisfaction regarding the working hours of the Library. It is found that the majority of the users from Aided College (47.5%), Self-Finance College (44.7%), Engineering Colleges (63.2%), Education Colleges (60.4%) and Pharmacy Colleges (55.3%) are satisfied with current working hours of the respective libraries. Yet a certain portion of users from Pharmacy College (20.2%) and Self-Financing College (12.6%) have said highly dissatisfied. More than (10.9%) of users from all colleges Libraries are found neutral. However, comparatively a good number of users from Self-financing colleges (16.4%), and Aided colleges (15.1%) are highly satisfied.

Overall analysis of all the Libraries discloses that most of the users (50.5%) responded as satisfied, even though (14.1%) users suggest that they are 'highly satisfied' with the present working hours. Moreover 12.3% users are at 'dissatisfied' and moreover (14.2%) of users are 'highly dissatisfied' with the working hours of the library. A small number of users (10.9%) are at the state of neutral with the working hour. The authorities will try to fulfill the requirement of the dissatisfied and highly dissatisfied groups instead of ignoring them. Chi square value 71.9 and P.value zero at 1% level of significance shows a significant variation in the satisfaction level of working hours among the users of various college libraries in Solapur University.

It is deduced that the level of satisfaction at the working hours of Library is observed more or less satisfactory among the users. Still the authorities should take essential steps to upgrade few libraries in Solapur University area up to the standard of state level libraries providing maximum services and extended working hours to their users.

## Conclusion

Colleges affiliated to Solapur University have very good infrastructure and well-developed libraries. The availability of resources and services for quality information in libraries do have an important influence on the satisfaction of users. They not only come back, but interact well of the library to other users when users are satisfied with library information resources. The libraries under the study also have certain deficiencies, like any institution. A good percentage of users do not know about ICT or OPAC services. Their library facilities, resources and services have to be strengthened. Advanced services must be provided to students and teachers by the libraries and they will get the optimum benefits.